\documentclass{ws-ijmpd}

\begin{document}

\def\nocropmarks{\vskip5pt\phantom{cropmarks}}

\renewcommand\ArtDir{./}

\let\trimmarks\nocropmarks      

\markboth{R. Ruffini, C.L. Bianco, P. Chardonnet, F. Fraschetti, V. Gurzadyan, S.-S. Xue}{On the instantaneous spectrum of gamma-ray bursts}

\catchline{}{}{}

\title{ON THE INSTANTANEOUS SPECTRUM OF GAMMA-RAY BURSTS}

\author{\footnotesize REMO RUFFINI, CARLO LUCIANO BIANCO and SHE-SHENG XUE}
\address{ICRA --- International Center for Relativistic Astrophysics and Dipartimento di Fisica,\\ Universit\`a di Roma ``La Sapienza'', Piazzale Aldo Moro 5, I-00185 Roma, Italy.}

\author{\footnotesize PASCAL CHARDONNET}
\address{ICRA --- International Center for Relativistic Astrophysics and Universit\'e de Savoie,\\ LAPTH - LAPP, BP 110, F­74941 Annecy-le-Vieux Cedex, France.}

\author{\footnotesize FEDERICO FRASCHETTI}
\address{ICRA --- International Center for Relativistic Astrophysics and Universit\`a di Trento,\\ Via Sommarive 14, I-38050 Povo (Trento), Italy.}
 
\author{\footnotesize VAHE GURZADYAN}
\address{ICRA --- International Center for Relativistic Astrophysics and Yerevan Physics Institute,\\ Alikhanian Brothers Street 2, 375036, Yerevan-36, Armenia.}

\maketitle

\pub{Received (received date)}{Revised (revised date)}

\begin{abstract}
A theoretical attempt to identify the physical process responsible for the afterglow emission of Gamma-Ray Bursts (GRBs) is presented, leading to the occurrence of thermal emission in the comoving frame of the shock wave giving rise to the bursts. The determination of the luminosities and spectra involves integration over an infinite number of Planckian spectra, weighted by appropriate relativistic transformations, each one corresponding to a different viewing angle in the past light cone of the observer. The relativistic transformations have been computed using the equations of motion of GRBs within our theory, giving special attention to the determination of the equitemporal surfaces. The only free parameter of the present theory is the ``effective emitting area'' in the shock wave front. A self consistent model for the observed hard-to-soft transition in GRBs is also presented. When applied to GRB~991216 a precise fit $\left(\chi^2\simeq 1.078\right)$ of the observed luminosity in the $2$--$10$ keV band is obtained. Similarly, detailed estimates of the observed luminosity in the $50$--$300$ keV and in the $10$--$50$ keV bands are obtained.
\end{abstract}

\keywords{black hole physics --- gamma rays: bursts --- gamma rays: observations --- gamma rays: theory ---  radiation mechanisms: thermal --- radiation mechanisms: general}

\section{Introduction}

Gamma-Ray Bursts (GRBs), following the observations by the BATSE instrument (Paciesas et al.\cite{batse4b}), have been characterized by a few global parameters (see e.g. Fishman \& Meegan\cite{fm95}) such as the fluence, the characteristic duration ($T_{90}$ or $T_{50}$), and the global spectral distribution given e.g. by the Band relation (Band et al.\cite{b93}). After the discovery of the afterglow (Costa et al.\cite{c97}), two additional important parameters have been added: the power-law indexes of the afterglow and the source luminosity.

It has become clear that a variety of different eras are present in the GRB data and that GRBs quite possibly have the most extreme time-variation of any phenomena in nature (see e.g. Ruffini et al.\cite{Brasile} and references therein). We present here an attempt to derive from first principles the instantaneous GRB luminosity in selected energy bands and the GRB spectra. We use GRB~991216 as the prototype (see Ruffini et al.\cite{rbcnr02}) since this source offers a superb set of data by BATSE in the $50$--$300$ keV band (see BATSE Rapid Burst Response\cite{brbr99}) and by R-XTE and Chandra in the $2$--$10$ keV band (see Piro et al.\cite{p00}, Corbet \& Smith\cite{cs00}) to be compared to the theoretically predicted ones in the $2$--$300$ keV range. We also give physical reasons for the often mentioned hard-to-soft transition observed in the majority of GRBs (see e.g. Frontera et al.\cite{fa00}, Ghirlanda et al.\cite{gcg02}, Piran\cite{p99}, Piro et al.\cite{p99b}).

\section{The model}

The complete dynamics of GRB 991216 has been computed (see Ruffini et al.\cite{Brasile}). The initial conditions we adopted for this source at $t=10^{-21}$ s $\sim 0$ s are a spherical shell of $e^+$-$e^-$-photon neutral plasma lying between the radii $r_0=6.03 \times 10^6$ cm and $r_1=2.35\times10^8$ cm: the temperature of such a plasma is $2.2$ MeV, the total energy $E_{tot}=4.83\times 10^{53}$ erg and the total number of pairs $N_{e^+e^-}=1.99\times 10^{58}$. These conditions have been derived from vacuum polarization processes occurring in the dyadosphere of an ElectroMagnetic Black Hole (EMBH) (Ruffini\cite{rukyoto}, Preparata et al.\cite{prx98}).

The optically thick electron-positron plasma self-propels itself outward reaching ultrarelativistic velocities (Ruffini et al.\cite{rswx99}), then interacts with the remnant of the progenitor star and by further expansion becomes optically thin (Ruffini et al.\cite{rswx00}). As the transparency condition is reached, the {\em Proper}-GRB (P-GRB) is emitted with an extremely relativistic shell of Accelerated Baryonic Matter (the ABM pulse, see Ruffini et al.\cite{lett2}). It is this ABM pulse which gives origin to the afterglow through its interaction with the ISM, whose average density is assumed to be $\left<n_{ism}\right> = 1$ particle/cm$^3$. In such a collision the ``fully radiative condition'' is implemented (see Ruffini et al.\cite{Brasile} for details): the internal energy $\Delta E_{\rm int}$ which results is instantaneously radiated away.

The equations of motion in our model depend only on two free parameters: the total energy $E_{tot}$, which coincides with the dyadosphere energy $E_{dya}$, and the amount $M_B$ of baryonic matter left over from the gravitational collapse of the progenitor star, which is determined by the dimensionless parameter $B=M_Bc^2/E_{dya}$. These two parameters have been determined by fitting, with high accuracy, the bolometric intensity and the slope of the afterglow (Ruffini et al.\cite{lett2}). We have also fit, again with high accuracy, the substructures observed in the E-APE which result from inhomogeneities in the ISM, still maintaining an average density distribution $\left<n_{ism}\right> = 1$ particle/cm$^3$ (Ruffini et al.\cite{rbcfx02_letter}).

\section{The newly assumed origin of the afterglow X- and $\gamma$-ray radiation}

Here we adopt three basic assumptions: a) the resulting radiation as viewed in the comoving frame during the afterglow phase has a thermal spectrum and b) the ISM swept up by the front of the shock wave, with a Lorentz gamma factor between $300$ and $2$, is responsible for this thermal emission. We also adopt, like in our previous papers (Ruffini et al.\cite{lett1,lett2,Brasile,rbcfx02_letter}), that c) the expansion occurs with spherical symmetry. These three assumptions are different from the ones adopted in the GRB literature, in which the afterglow emission is believed to originate from synchrotron emission in the production of the shock or reverse shock generated when the assumed jet-like ejecta encounter the external medium (see e.g. Giblin et al.\cite{gca02} and references therein).

The structure of the shock is determined by mass, momentum and energy conservation, i.e., the constancy of the specific enthalpy, which are standard conditions in shock rest frames (Zel'dovich \& Rayzer\cite{zr66}) and have already been used in our derivation (Ruffini et al.\cite{Brasile}). The only additional free parameter of our model is the size of the ``effective emitting area'' in the shock wave front: $A_{eff}$.

The temperature $T$ of the black body in the co-moving frame is then
\begin{equation}
T=\left(\frac{\Delta E_{\rm int}}{4\pi r^2 \Delta \tau \sigma {\cal R}}\right)^{1/4}\, ,
\label{tcom}
\end{equation}
where
\begin{equation}
{\cal R}=\frac{A_{eff}}{A_{abm}}
\label{rdef}
\end{equation}
is the ratio between the ``effective emitting area'' and the ABM pulse surface $A_{abm}$, $\Delta E_\mathrm{int}$ is the internal energy developed in the collision with the ISM in a time interval $\Delta \tau$ in the co-moving frame and $\sigma$ is the Stefan-Boltzmann constant. The ratio ${\cal R}$, which is a priori a function that varies as the system evolves, is evaluated at every given value of the laboratory time $t$.

All the subsequent steps are now uniquely determined by the equations of motion of the system. The basic tool in this calculation involves the definition of the EQuiTemporal Surfaces (EQTS) for the relativistic expanding ABM pulse as seen by an asymptotic observer. See Fig.~1 in Ruffini et al.\cite{rbcfx02_letter} and Bianco \& Ruffini\cite{eqts_apj,eqts}.

We are now ready to evaluate the source luminosity in a given energy band. The source luminosity at a detector arrival time $t_a^d$, per unit solid angle $d\Omega$ and in the energy band $\left[\nu_1,\nu_2\right]$ is given by (see Ruffini et al.\cite{Brasile}):
\begin{equation}
\frac{dE_\gamma^{\left[\nu_1,\nu_2\right]}}{dt_a^d d \Omega } = \int_{EQTS} \frac{\Delta \varepsilon}{4 \pi} \; v \; \cos \vartheta \; \Lambda^{-4} \; \frac{dt}{dt_a^d} W\left(\nu_1,\nu_2,T_{arr}\right) d \Sigma\, ,
\label{fluxarrnu}
\end{equation}
where $\Delta \varepsilon=\Delta E_{int}/V$ is the energy density released in the interaction of the ABM pulse with the ISM inhomogeneities measured in the comoving frame, $\Lambda=\gamma(1-(v/c)\cos\vartheta)$ is the Doppler factor, $W\left(\nu_1,\nu_2,T_{arr}\right)$ is an ``effective weight'' required to evaluate only the contributions in the energy band $\left[\nu_1,\nu_2\right]$, $d\Sigma$ is the surface element of the EQTS at detector arrival time $t_a^d$ on which the integration is performed (see also Ruffini et al.\cite{rbcfx02_letter}) and $T_{arr}$ is the observed temperature of the radiation emitted from $d\Sigma$:
\begin{equation}
T_{arr}=\frac{T}{\gamma \left(1-\frac{v}{c}cos\vartheta\right)}\frac{1}{(1+z)}\, .
\label{Tarr}
\end{equation}

The ``effective weight'' $W\left(\nu_1,\nu_2,T_{arr}\right)$ is given by the ratio of the integral over the given energy band of a Planckian distribution at a temperature $T_{arr}$ to the total integral $aT_{arr}^4$:
\begin{equation}
W\left(\nu_1,\nu_2,T_{arr}\right)=\frac{1}{aT_{arr}^4}\int_{\nu_1}^{\nu_2}\rho\left(T_{arr},\nu\right)d\left(\frac{h\nu}{c}\right)^3\, ,
\label{effweig}
\end{equation}
where $\rho\left(T_{arr},\nu\right)$ is the Planckian distribution at temperature $T_{arr}$:
\begin{equation}
\rho\left(T_{arr},\nu\right)=\frac{2}{h^3}\frac{h\nu}{\exp^{h\nu/\left(kT_{arr}\right)}-1}
\label{rhodef}
\end{equation}

\section{The best fit of the observed flux in selected energy bands}

\begin{figure}[htbp]
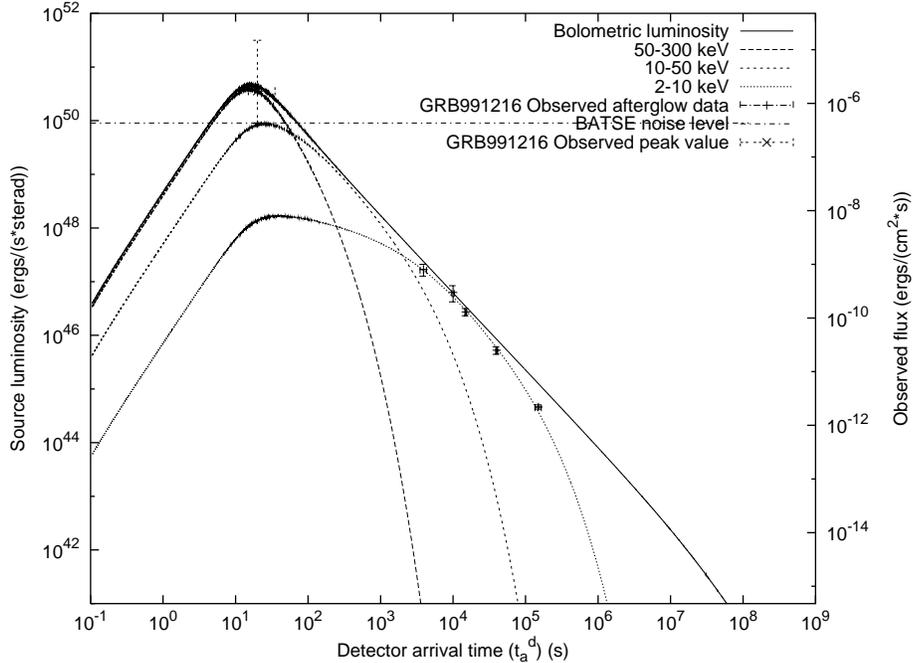

\figurebox{}{}{spectfit}
\caption{Best fit of the afterglow data of GRB~991216. The solid curve is the bolometric luminosity. See Ruffini et al.$^{18}$ for the radial approximation and Ruffini et al.$^{5,19}$ for the relativistic analysis of the off-axis contributions. The three dotted curves correspond to the luminosities in the bands $50$--$300$ keV, $10$--$50$ keV and $2$--$10$ keV respectively. Near the E-APE, where the BATSE data are present, almost all the luminosity is in the $50$--$300$ keV band. The afterglow data from R-XTE and Chandra (see Halpern et al.$^{25}$) in the $2$--$10$ keV are perfectly fit by the corresponding luminosity curve (see also Fig.~\ref{spectfit_zoom}).}
\label{spectfit}
\end{figure}

\begin{figure}[htbp]
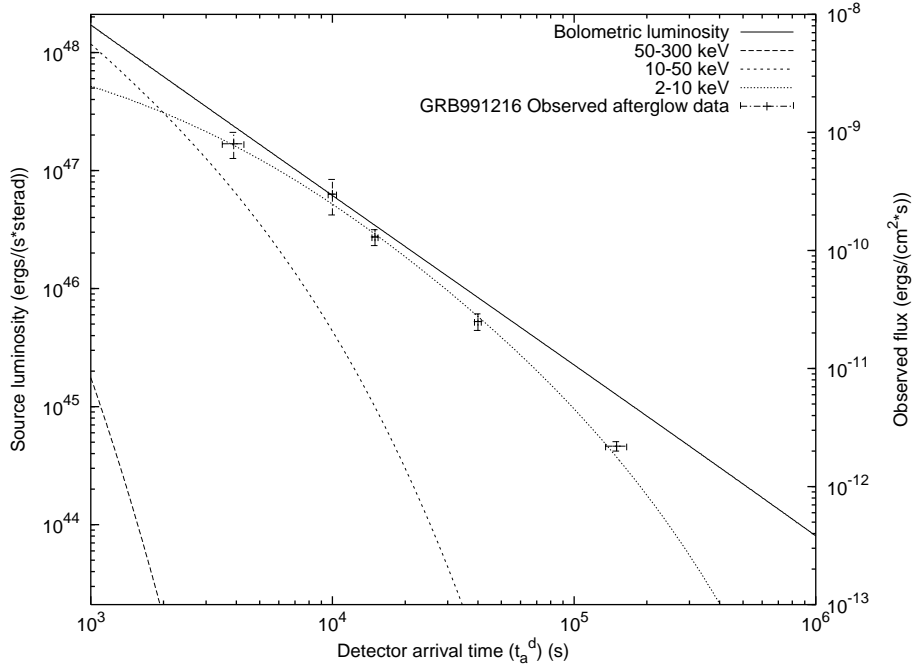

\figurebox{}{}{spectfit_zoom}
\caption{This is an enlargement of Fig.~\ref{spectfit} in the region of the afterglow data in the $2$--$10$ keV band from the R-XTE and Chandra satellites, showing the perfect agreement between the theoretical curve and the observational data. The reduced $\chi^2$ value for this fit is $\chi^2\simeq 1.078$.}
\label{spectfit_zoom}
\end{figure}

We can now proceed to the best fit of the observed data using GRB~991216 as the prototype. Such an estimate is perfectly well defined from a theoretical point of view, although from a numerical point of view the integration on all the EQTS and the associated relativistic transformations have raised unprecedented and time consuming difficulties. Almost $10^8$ paths with different temperatures and different Lorentz boosts had to be considered in the integration over the EQTS. We give in Figs.~\ref{spectfit}--\ref{spectfit_zoom} the results for the three energy bands $50$--$300$ keV (BATSE), $2$--$10$ keV (R-XTE, Chandra) and $10$--$50$ keV. It is most remarkable that the best fit is obtained simply by a factor ${\cal R}$, which is monotonically varying in the range:
\begin{equation}
3.01\times 10^{-8} \ge {\cal R} \ge 5.01 \times 10^{-12}
\label{Rval}
\end{equation}
respectively in correspondence with the beginning of the afterglow emission and the last observation by Chandra at $\sim 37$ hr after the GRB. We point out the perfect agreement with the data obtained by BATSE Rapid Burst Response\cite{brbr99} in the energy range $50$--$300$ keV  (see dashed line in Fig.~\ref{spectfit}). Very impressive is the fit of the data obtained by the R-XTE and Chandra satellites (see Halpern et al.\cite{ha00}) in the energy range $2$--$10$ keV (see dotted line in Figs.~\ref{spectfit}--\ref{spectfit_zoom}). These data are fitted with a $\chi^2\simeq 1.078$. This fit covers a time span of $\sim 10^6$ s and is impressive if we recall that it is a function of the single parameter ${\cal R}$. The fit can be further improved, reaching a $\chi^2\simeq 0.48$, when a radial dependence in $\left<n_{ism}\right>$ is introduced, ranging from $\left<n_{ism}\right>\simeq 1$ particle/cm$^3$ in the E-APE region ($r\simeq 5\times 10^{16}$ cm) to $\left<n_{ism}\right>\simeq 3$ particle/cm$^3$ in the latest afterglow phases ($r\simeq 4\times 10^{17}$ cm). Both in Fig.~\ref{spectfit} and Fig.~\ref{spectfit_zoom} the solid line gives the bolometric luminosity (see details in Ruffini et al.\cite{Brasile}).

\section{On the time integrated spectra and the hard-to-soft spectral transition}

\begin{figure}[htbp]
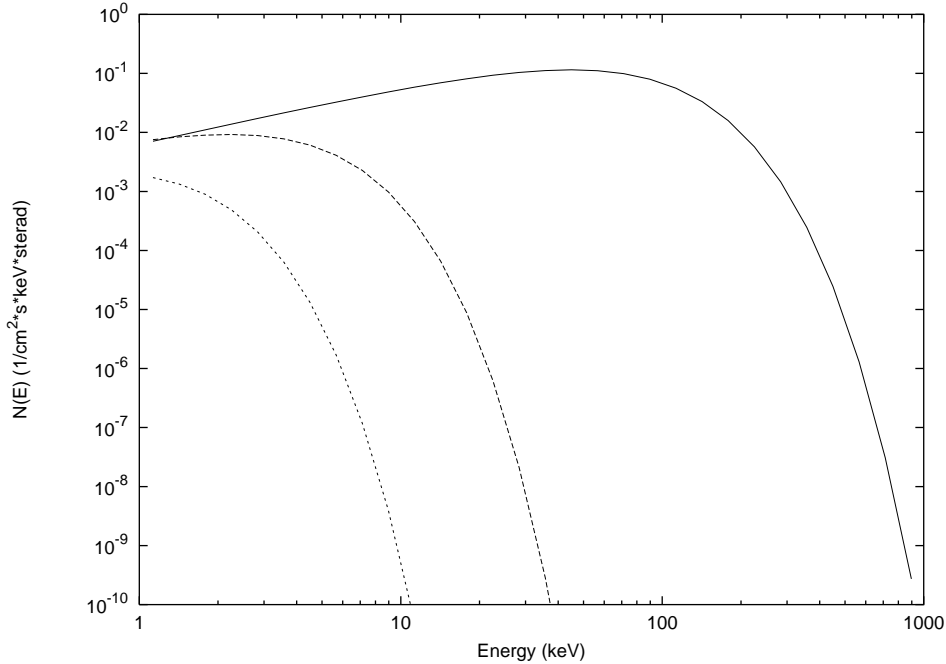

\figurebox{}{}{spectrum}
\caption{The instantaneous spectra of the radiation observed in GRB~991216 at three different EQTS respectively, from top to bottom, for $t_a^d=10$ s, $t_a^d=10^4$ s and $t_a^d=1.45\times10^5$ s. These diagrams have been computed assuming a constant $\left<n_{ism}\right>\simeq 1$ particle/cm$^3$ and clearly explains the often quoted hard-to-soft spectral evolution in GRBs.}
\label{spectrum}
\end{figure}

We turn now to the much debated issue of the origin of the observed hard-to-soft spectral transition during the GRB observations (see e.g. Frontera et al.\cite{fa00}, Ghirlanda et al.\cite{gcg02}, Piran\cite{p99}, Piro et al.\cite{p99b}). We consider the instantaneous spectral distribution of the observed radiation for three different EQTS:
\begin{itemize}
\item $t_a^d=10$ s, in the early radiation phase near the peak of the luminosity,
\item $t_a^d=1.45\times 10^5$ s, in the last observation of the afterglow by the Chandra satellite, and
\item $t_a^d=10^4$ s, chosen in between the other two (see Fig.~\ref{spectrum}).
\end{itemize}
The observed hard-to-soft spectral transition is then explained and traced back to:
\begin{enumerate}
\item a time decreasing temperature of the thermal spectrum measured in the comoving frame,
\item the GRB equations of motion,
\item the corresponding infinite set of relativistic transformations.
\end{enumerate}
A clear signature of our model is the existence of a common low-energy behavior of the instantaneous spectrum represented by a power-law with index $\alpha = +0.9$. This prediction will be possibly verified in future observations.

\begin{figure}[htbp]
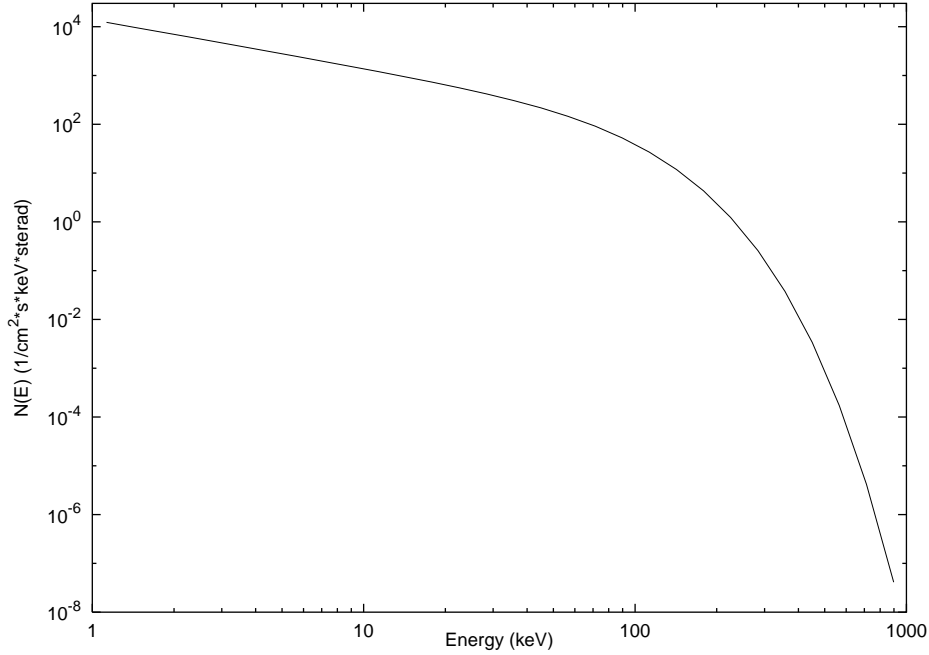

\figurebox{}{}{spectband}
\caption{The time-integrated spectrum of the radiation observed in GRB~991216. The low energy part of the curve below $10$ keV is fit by a power-law with index $\alpha = -1.05$ and the high energy part above $500$ keV is fit by a power-law with an index $\beta < -16$.}
\label{spectband}
\end{figure}

Starting from these instantaneous values, we integrate the spectra in arrival time obtaining what is usually fit in the literature by the ``Band relation'' (Band et al.\cite{b93}). Indeed we find for our integrated spectra a low energy spectral index $\alpha=-1.05$ and an high energy spectral index $\beta < -16$ when interpreted within the framework of a Band relation (see Fig.~\ref{spectband}). This theoretical result can be submitted to a direct confrontation with the observations of GRB 991216 and, most important, the entire theoretical framework which we have developed can now be applied to any GRB source. The so obtained theoretical predictions on the luminosity in fixed energy bands can be then straightforwardly confronted with the observational data.

\section{Conclusions}

In addition to the above results, we have also applied our model to GRB~980425, which is one of the weakest GRBs observed, with an energy of the order of $\sim 10^{48}$ ergs (see Ruffini et al.\cite{cospar02,r03mg10}, Ruffini\cite{r03tokyo}). Our model then applies over a range of energies spanning 6 orders of magnitude.

The fundamental novel point here is the assumption of the thermal origin of the X and $\gamma$ radiation of the afterglow in the comoving frame of the shock front. The fit of the data in Figs.~\ref{spectfit}--\ref{spectfit_zoom} gives a most clear and unambiguous support from the observations to this theoretical approach.

All the works in the current literature tries to explain the afterglow emission by a very complex process implying magnetic fields, jet-like ejecta, emission by a forward shock and a reverse shock (see e.g. Piran\cite{p99}, van Paradijs et al.\cite{vpkw00}, M\'esz\'aros\cite{m02} and references therein). In our approach we evidence the existence of a much simpler process, directed forward, basically spherically symmetric and originating by a simple thermal emission in the comoving frame of the shock.

We are grateful to an anonymous referee for pointing out that Blinnikov et al.\cite{ba99} did argue that nonthermally looking GRB spectra can indeed be formed by a superposition of a set of thermal black body spectra with a temporal power-law evolution of the temperature. In our treatment not only time but also space integration on the EQTS takes place. This effect was explicitly omitted in the interesting paper of Blinnikov et al.\cite{ba99}: while their instantaneous GRB spectra are thermal, in our approach each instantaneous spectrum is derived from an infinite set of foliations of events on the EQTS, each one characterized by a different thermal spectrum in the comoving frame boosted by a different relativistic transformation obtained from the EOM.

We emphasize that these results are extremely sensitive to the structure of the EQTS and to the theoretical assumptions adopted for each GRB era (see examples in Ruffini et al.\cite{Brasile}, Bianco \& Ruffini\cite{eqts,eqts_apj}). Due to the enormous redundancy built into the almost $10^8$ different paths mentioned above, possibly unprecedented in physics and astrophysics, we can assert the uniqueness of the solution. We also conclude that there is a marked difference (see Fig.~\ref{spectfit}) between the bolometric intensity of the afterglow, with a simple power-law behavior with an index $n=-1.6$ in the decreasing part, and the actual luminosity in a fixed bandwidth, which can have a complex dependence on time. Such a complex behavior could be erroneously interpreted as a broken power-law supporting the existence of jet-like structures in GRBs.

The physical reasons justifying the assumptions in Eqs.(\ref{tcom}--\ref{rdef}) are presented in Ruffini et al.\cite{Spectr2}.

\end{document}